# Drag Reduction Performance of Mechanically Degraded Dilute Polyethylene Oxide Solutions

Yasaman Farsiani[1], Zeeshan Saeed[2] and Brian R. Elbing[3,a]

[1]Mechanical & Aerospace Engineering, Oklahoma State University,
201 GAB, Stillwater, OK 74078; yasaman.farsiani@okstate.edu

[2]Mechanical & Aerospace Engineering, Oklahoma State University,
201 GAB, Stillwater, OK 74078; zeeshan.saeed@okstate.edu

[3]Mechanical & Aerospace Engineering, Oklahoma State University,
201 GAB, Stillwater, OK 74078; elbing@okstate.edu

## Abstract

Mechanical degradation of dilute solutions of polyethylene oxide (PEO) via chain scission was investigated within a turbulent pipe flow. Comparisons of the drag reduction performance with and without degradation were made by matching the onset of drag reduction conditions, which has been shown for PEO to be related to the mean molecular weight. The bulk flow behavior of both the degraded and non-degraded samples were generally consistent with trends observed in the literature, but a subset of conditions showed significant deviation in the slope increment (drag reduction performance) between the degraded and non-degraded samples. When they deviated, the degraded samples were consistently more efficient than the non-degraded samples even though they had the same mean molecular weight. The deviations were shown to scale with the normalized difference between the initial and final molecular weights. The current data and analysis as well as the literature suggests that the deviations in the polymer performance (slope increment) are related to changes in the molecular weight distribution. More specifically, the improved performance of the degraded samples relative to the non-degraded ones at the mean molecular weight of the degraded sample indicates an excess of longer polymer chains since the higher chain fractions in a degraded solution more effectively control the flow properties when within a certain degree of degradation and Reynolds number.

**Keywords:** polymer drag reduction; degradation; pipe flow; pressure drop; experimental

[a] Author to whom correspondence should be addressed. Electronic mail: elbing@okstate.edu

# 1 Introduction

The ability to reduce skin-friction with polymer solutions, historically referred to as Toms effect, has been known since the late 1940s [1-3]. Since that point, there have been numerous successful studies that have produced various applications [4] with most associated with internal flows [5-6]. Most active research in polymer drag reduction (PDR) focuses on developing a fundamental understanding of PDR to enable external flow applications such as marine vehicles [7-13], which in 2010 there was a successful application of PDR to improve ship speed on a sailing vessel [14]. One of the primary limiting factors for advancing PDR to external flows (as well as expanding internal flow applications) is polymer degradation. Polymer degradation is known to be dependent on many initiating factors such as oxidative and bacterial action, free radical interaction (chemical degradation), thermal degradation and mechanical degradation [15-20]. The aging of polymer solution has also been found to contribute to polymer solution degradation [21]. The current study, however, focuses solely on degradation due to chain scission induced from turbulent flow structure-polymer chain interaction (i.e., mechanical degradation in a turbulent pipe flow). Many theoretical/computational models detailing the mechanics of the physical process of polymer-chain degradation based on their interaction with the basic turbulent flow structures (e.g. horseshoe/hairpin vortices) are available [22-24].

While the literature for polymer degradation (mostly for internal flows) is vast, a brief review of key papers that influence the current work is provided here for completeness. The flow induced shear (mechanical force) on the polymer chain can be generated from abrupt changes in flow geometry (e.g. pumps, valves and perforations [25]) or large mean velocity gradients such as those experienced at the wall of high Reynolds number turbulent flows [11,26-28]. Initial studies (e.g. [29]) discovered that it was extremely challenging to produce a setup that could isolate the degradation to the flow region of interest. Even more recent studies have frequently concluded that the majority of the degradation was produced at the entrance to their test facility [19,28,30].

Paterson and Abernathy [31] was one of the earliest investigations on the impact of flow-assisted (mechanical) degradation on the molecular weight distribution (e.g. polydispersity), which it concluded that degradation influences the resulting flow properties. Yu et al. [32] used monodisperse polystyrene and polydisperse polyisobutene samples in oils under high shear rates to show that the molecular weight distribution narrowed for the polydispersed samples and only had a slight broadening of the monodispersed sample. More specifically, the distributions revealed that the breaking of the chains was not a random process. Hinch [33] developed a formula estimating the required force to break a molecular chain at a given location within the chain. This theoretical evaluation showed that the maximum strain developed in stretching the chain was at its center. Subsequently, Horn and Merrill [34] showed that turbulence induced scission of macromolecules in dilute solutions preferentially break at the midpoint of the chain. Moreover, Odell et al. [35] studied extensional flows produced in cross slot devices with low molecular weight ($M_w < 10^6$) polyethylene oxide (PEO), which was at sufficiently low molecular weights for measurements of the molecular weight distribution. These results showed that the molecular weight distribution had another peak in addition to the original one at half the molecular weight, indicating scission of the chain at its midpoint. More recently, a simulation of flow-induced polymer chain scission [36], validated the midpoint scission hypothesis under the condition that the elongation rate was comparable to the critical elongation rate, then the instantaneous segmental tensions attains a maximum at the chain midpoint. This has the consequence of the resulting daughter chains having a rather narrow distribution. However, it was also demonstrated that when the elongational rate is much larger than the critical elongational rate, scission can occur in the partially coiled chains resulting in scission occurring farther from the midpoint. This likely has a significant impact within wall-bounded turbulent flows where the stress distribution varies significantly from the maximum at the wall to very weak away from the wall (e.g. pipe centerline or outside of a turbulent boundary layer).

      Hunston and Zakin [37] used turbulent drag reduction (similar to current study), viscosity and gel permeation chromatography (GPC) on polystyrene samples to assess the influence of concentration, molecular weight and molecular weight distribution on flow-assisted (mechanical) degradation. This showed that the onset of drag reduction provided information about the largest molecules in the flow while the flow rate dependence was related to the shape of the top part of the molecular weight distribution. Gampert and Wagner [38] used laboratory synthesized straight molecular chain polyacrylamide (PAM) in aqueous solutions to investigate the influence of molecular weight and polydispersity on drag-reducing effectiveness. Gampert and Wagner [38] artificially created polydispersity by mixing the high and low synthesized molecular weights in a single solution. This work made several conclusions that are consistent with available literature. Primarily, that the long chain molecules are pivotal in determining the flow properties of a solution due to their preferential mode of extension and hence degradation, when the difference in size of existing chains in a solution is reasonably high within a suitable range of Reynolds number.

      Historically, mechanical degradation has significantly limited the viable applications for PDR since polymers are generally more efficient at reducing drag (i.e. require lower concentrations to achieve a desired drag reduction) the longer the polymer chain (i.e. higher molecular weight), but the longer the polymer chain the more susceptible it is to chain scission. Internal flows have typically avoided this problem by using stiffer polymers (e.g. PAM), and some applications have shown an increase in resistance to mechanical degradation with increasing concentration [39]. But

the use of commercial grade PAM is not suitable for investigations involving the influence of molecular weight on drag reduction and mechanical degradation because they have a branched chain formation and the presence of copolymers [38]. Instead, PDR studies have focused primarily on high molecular weight PEO, as has been the case in external flow studies, because PEO has the ability to achieve maximum drag reduction (MDR) with concentrations of ~10 parts-per-million (ppm). In addition, PEO avoids the rheological issues faced with commercial grade PAM. The high efficiency of PEO is ideal for external flows, which continuously dilute the injected polymer solution into the developing boundary layer. However, as a result, polymer degradation has had a significant impact on PDR external applications and even the ability to study PDR within turbulent boundary layers.

Elbing et al. [11] developed a fundamental scaling law for the evolution of the mean molecular weight within a developing high-shear turbulent boundary layer, which requires an estimate of the steady-state molecular weight for the given local shear rate. This was produced from the universal scaling law for chain scission [40], given that the nominal bond strengths for carbon-carbon and carbon-oxygen bonds are 4.1 nN and 4.3 nN [41], respectively. In light of this observation, a review of literature that has reported PDR modifications to the near-wall velocity profile of a turbulent boundary layer with PEO [12,26,27,42-44], shows that many of the reported conditions [12,26,27] experienced significant changes in the mean molecular weight between the injection and measurement locations even though this was not accounted for in their analysis. This is particularly problematic when studying high drag reduction (HDR; >40% drag reduction), which recent computational [45] and experimental [12,46] work (see Figure 1; where the streamwise velocity and wall-normal distance are scaled with inner variables) has shown that modifications to the near-wall velocity profile deviate from the classical view that assumes the near-wall momentum distribution is independent of polymer properties. Elbing et al. [12] showed that Reynolds number was insufficient to collapse the available experimental turbulent boundary layer data, which suggests that the remaining scatter in the results must be related to polymer properties. These polymer properties are sensitive to the molecular weight, which means that in addition to an evolving polymer concentration distribution, there is also an evolving molecular weight distribution that needs to be accounted for to properly study HDR in a turbulent boundary layer. Recently this has motivated an alternative approach, which is to develop a polymer ocean at a uniform concentration that has been mechanically degraded to a steady-state molecular weight [47,48]. Then the developing boundary layer would have a known and uniform polymer concentration and molecular weight. However, this requires a proper understanding of the impact of mechanical degradation via chain scission on the drag reduction performance of PEO, which is the focus of the current study. Specifically, the current study prepares degraded and non-degraded samples at the same mean molecular weight and does a comparative analysis of their drag reduction performance.

## 2 Experimental Method

### *2.1 Polymer preparation*

PEO was the only polymer used in this study, which has a structural unit (monomer) of (-O-CH$_2$-CH$_2$-) that results in a polymer backbone consisting of carbon-carbon (C-C) and carbon-oxygen (C-O) bonds. Five molecular weights of PEO were tested with manufacturer specified mean molecular weights of 0.6 1.0, 2.0 and 5.0×10$^6$ g/mol (Sigma Aldrich) and 4.0×10$^6$ g/mol (WSR301, Dow chemical). As previously stated, these high molecular weight polymers are highly susceptible to mechanical degradation within shear flows. Degradation depends on the molecular weight, polymer concentration, solvent type, turbulent intensity and flow geometry [49,50]. Polymer solutions were prepared by sprinkling dry powder into a water jet prior to contacting the free surface to avoid formation of polymer aggregates. Chlorine in the solvent (water) can cause polymer degradation [51,52]. Therefore, when solutions required hydration longer than 12 hours the background chlorine was removed by adding trace amount of sodium thiosulfate, which residual sodium thiosulfate and the resultant products of the reaction with chlorine has been shown to not impact the PDR performance [52]. Stock solutions were prepared at relatively high concentrations (1000-5000 ppm), which once fully hydrated, additional water was added to dilute the sample to the desired test concentration.

Polymer solution concentrations are broadly categorized as dilute, semi-dilute and concentrated. In the dilute regime, each polymer chain is sufficiently distant from other polymer chains such that there is minimal interaction between chains. As the concentration increases, polymer chains eventually begin to overlap and become entangled, which these interactions alter the polymer properties. As the concentration further increases from semi-dilute to the concentrated regime, molecules cannot move freely and significant interpenetration occurs due to the lack of space. These changes are identified from their rheological properties [53]. The critical overlap concentrations $C^*$ and $C^{**}$ define the transition points from dilute to semi-dilute and semi-dilute to concentrated regimes, respectively. The first overlap concentration can be found from the inverse of the intrinsic viscosity ($[\eta]_0$), $C^* = [\eta]_0^{-1}$, which $[\eta]_0$ can be estimated from the Hark-Houwink relationship [54], $[\eta]_0 = 0.0125 M_w^{0.78}$. Table 1 provides the range of molecular weights and concentrations tested for degraded and non-degraded samples in this study as well as the corresponding intrinsic viscosity and $C^*$. The first overlap concentration was well above the test range at each molecular weight and, consequently, all testing was with dilute solutions.

## *2.2 Test facility and instrumentation*

The primary test facility was a pressure drop apparatus that was used to characterize polymer properties and acquire the gross flow behavior. A schematic of the setup is shown in Figure 2, including the pipe as well as the instrumentation. Test samples were placed in an 18.9-liter 316L stainless steel pressure vessel (740560, Advantc), which was sealed and pressurized to ~275 kPa during testing. A dip tube drew the polymer sample from the bottom of the pressure vessel and then pushed it into the pipe flow portion. It consisted of a 10.9 mm inner diameter (*d*) instrument grade seamless 316 stainless steel pipe (SS-T8-S-035-20, Swagelok) that was divided into 3 sections; the entrance length that was 150*d* long to achieve fully developed turbulent pipe flow, a 1.05 m long test section and the end (exit) length that had a V-shaped, 4.8 mm orifice needle valve at the outlet to control flowrate (this valve was also used as the primary means to accelerate the degradation of the polymer solutions). Given the valve properties and the current operation range, the maximum flow coefficient was nominally 0.22. The pressure drop across the test section was acquired at various Reynolds numbers with a differential pressure transducer (PX2300-5DI, Omega Engineering). The mass flowrate, and ultimately the average velocity within the pipe, was determined by measuring the mass of a sample on a 35 kg digital balance (CPWplus-35, Adam Equipment) while simultaneously recording the fill time with a stopwatch (RS-013, ProCoach). A more detailed discussion of the setup, instrumentation and uncertainty quantification is provided in Lander [55].

Polymer solution temperature was measured with a thermometer (25-125°F, TEL-TRU) and was held relatively constant throughout testing at 21±0.4 °C with a corresponding mean density ($\rho$) and kinematic viscosity ($\nu$) of 998 kg/m$^3$ and 1.0×10$^{-6}$ m$^2$/s, respectively. The pipe diameter and pressure drop length had uncertainties below 1%. Given the high accuracy of the pressure transducer (±0.25%), the largest uncertainty in the differential pressure was due to the experimental setup. Specifically, the holes (pressure taps) in the pipe walls required for the measurements were the largest source of error. Corrections were applied following the approach outlined in McKeon [56] with the corresponding uncertainty of ~3% for the pressure measurement. The mass flowrate was determined from measuring the fill time and mass during data collection. There are several sources of errors that were considered [55], but the largest source of error was the limitation of human reaction times. This resulted in uncertainties as large as 3%, but by increasing the measurement period (especially for low flow rate conditions) this uncertainty was reduced.

Propagating all sources of uncertainty results in uncertainties that significantly vary with flow condition resulting in Prandtl-von Kármán (P-K) coordinates $Re\sqrt{f}$ and $1/\sqrt{f}$ having typical uncertainties of ~6%. However, at low flow rates (i.e. at and below the onset of drag reduction) the uncertainty increases rapidly to well above 10% [55]. For these reasons measurements at the onset of drag reduction were not attempted, but rather measurements at higher flow rates with uncertainty below 10% were curve fitted and then extrapolated back to the onset condition. Subsequently, the analysis focuses on the variations of the slopes of these curve fits, so a more detailed uncertainty analysis on the impact of these uncertainties in the P-K coordinates on the slope and intercept of the curve fits was performed. Here error was introduced to each of the P-K coordinates such that a logarithmic curve fit takes the form of $1/\left(\sqrt{f+\epsilon_o}\right) = C_o \ln\left(Re\sqrt{f} + \epsilon_1\right) + C_1$, where $C_o, C_1$ are constants and $\epsilon_o, \epsilon_1$ are the uncertainties in $f$ and $Re\sqrt{f}$, respectively. Some algebraic manipulation and expanding the resulting expression in a binomial form results in

$$\frac{1}{\sqrt{f}}\left(1 - \frac{\epsilon_o}{2f} + \cdots\right) = C_o \ln(Re\sqrt{f}) + C_o \ln\left(1 + \frac{\epsilon_1}{Re\sqrt{f}}\right) + C_1.$$

Neglecting higher order terms and some addition rearranging reduces the relationship to

$$\frac{1}{\sqrt{f}} = C_o \ln(Re\sqrt{f}) + C_o \ln\left(1 + \frac{\epsilon_1}{Re\sqrt{f}}\right) + C_1 + \frac{\epsilon_o}{2f^{1.5}}.$$

Treating the error sources as being nominally constants, this shows that the uncertainty impacts the intercept more than the slope when $Re\sqrt{f}$ is large (current data $Re\sqrt{f} \sim 10^3$).

Like many previous degradation studies [19,28,37,49,57], the current work utilizes the drag reduction performance (determined from its behavior in this pressure drop apparatus) to investigate degradation. GPC and other such methods are preferred since they provide direct measurements of the molecular weight distribution, but GPC has proven to be impractical for high molecular weight PEO due to significant uncertainties in the analysis. One of the main issues that complicates this approach is that while PEO is soluble in tetrahydrofuran (THF), a common eluent for GPC, it is isorefractive with THF. This makes it so that it cannot be seen in that solvent with either index of refraction or light scattering detectors [14]. Another significant limitation in the use of GPC for estimating the polydispersity comes from its incapability to resolve low molecular weight fractions for PEO molecular weights as low as $2.5 \times 10^5$ g/mol [58]. Such lower molecular weight fractions are known to significantly affect the number average molecular weight of a sample necessary to evaluate polydispersity [38,58]. This renders the use of GPC for high molecular weight samples ineffective and so has not been made use of in this investigation. Thus, the current study quantifies the impact of mechanical degradation via chain scission on the drag reduction performance of PEO primarily from the resulting variations in the turbulent drag reduction performance. Based on previous studies using other polymer solutions, the likely impact on the molecular weight distribution is inferred.

## 3 Results

### 3.1 Non-degraded bulk flow characterization

It is common to present pipe flow skin-friction results in either a Moody diagram (Darcy-Weisbach friction factor versus Reynolds number) or in P-K coordinates ($f^{-1/2}$ versus $Re_d f^{1/2}$). For the current study, P-K coordinates are used, where $Re_d$ ($= \rho V d/\mu$) is pipe diameter-based Reynolds number, $f$ is the Fanning friction factor ($f = 2\tau/\rho V^2$), $\rho$ is the fluid density, $V$ is the mean velocity, $d$ is the pipe diameter, $\mu$ is the fluid dynamic viscosity and $\tau$ is the wall shear stress. Assuming fully developed pipe flow, the wall shear stress ($\tau$) is directly related to the pressure drop across a given length of pipe ($\tau = \Delta p\, d/4\, \Delta x$), where $\Delta p$ is the pressure drop measured over the pipe length $\Delta x$. The physical significance of P-K plots is that the ordinate is the ratio of the bulk fluid velocity to the turbulent friction velocity (divided by $\sqrt{2}$) and the abscissa is the ratio of the pipe diameter (outer length scale) to the viscous wall unit (inner length scale) (multiplied by $\sqrt{2}$). The skin-friction curve for Newtonian turbulent pipe flow in these coordinates is well represented by the P-K law, Eq. (1). Newtonian (water) results from the current setup are included in Figure 3. These results are well approximated by the P-K law, which is also included for comparison.

$$\frac{1}{\sqrt{f}} = 4.0\, log_{10}(Re_d \sqrt{f}) - 0.4 \tag{1}$$

With the addition of drag reducing polymer (PEO) solution, the results are shifted above the P-K law. The amount of increase is limited by the empirically derived maximum drag reduction (MDR) asymptote [59] given in Eq. (2). The current study focuses on results within the polymeric region, which is at intermediate drag reduction levels between the MDR asymptote and the P-K law. The data within the polymeric regime are fitted following the form given in Eq. (3) [60]. Here $\delta$ is the slope increment and $W^*$ is the onset wave number, which both are dependent on the polymer properties. Furthermore, the slope increment ($\delta$) is the change in slope relative to the P-K law slope, and the onset wave number ($W^*$) can be shown to be equal to the reciprocal of the viscous wall unit at the onset of drag reduction. The onset of drag reduction is identified by the intersection of the P-K law and the polymeric data fitted with Eq. (3). Note that below this minimum shear rate required to initiate drag reduction, the polymer solutions follow the P-K law, which is indicative of the need for a sufficient amount of shear to stretch the polymer chains and active the drag reduction mechanism [61,62]. The onset of drag reduction for a given polymer type and molecular weight has been shown to have a negligible dependence on the concentration [30,60]. Current polymeric results using PEO at a $M_w = 2 \times 10^6$ g/mol and at concentrations from 100 to 500 ppm are also provided in Figure 3. These results show that the slope increment increases with increasing polymer concentration ($C$) while the onset of drag reduction (intersection of P-K law and polymeric data fit) remains nearly constant for all three samples tested.

$$\frac{1}{\sqrt{f}} = 19.0 \, log_{10}(Re_d\sqrt{f}) - 32.4 \qquad (2)$$

$$\frac{1}{\sqrt{f}} = (4.0 + \delta)log_{10}(Re_d\sqrt{f}) - 0.4 - \delta log_{10}(2dW^*) \qquad (3)$$

While the onset of drag reduction remains constant for a given molecular weight, it is sensitive to the mean molecular weight. Generally, the higher the $M_w$ the lower the Reynolds number at the onset of drag reduction. Vanapalli et al. [30] compiled PEO data [60] to establish an empirical relationship between the onset of drag reduction shear rate ($\gamma^*$) and the mean molecular weight ($M_w$), $\gamma^* = 3.35 \times 10^9/M_w$. This allows for the mean molecular weight to be determined if the wall shear rate at the onset of drag reduction is known. The wall shear rate at the onset of drag reduction is determined by calculating the intersection between the polymeric best-fit curve and the P-K law. The intersection provides the corresponding onset of drag reduction Fanning friction factor ($f^*$) and the onset of drag reduction Reynolds number ($Re_d^*$). Given the definition of the Fanning friction factor and the relationship between shear stress and shear rate at the wall ($\gamma = \tau/\rho\nu$), the onset shear rate at the onset of drag reduction can be determined from $f^*$, $\gamma^* = V^2 f^*/2\nu$. Thus, the mean molecular weight of the PEO polymer solutions can be inferred from the P-K plots. Table 2 provides a summary of the non-degraded conditions tested, including mean molecular weight ($M_w$), the resulting slope increment ($\delta$), onset wave number ($W^*$) and the shear rate at the onset of drag reduction ($\gamma^*$). These results demonstrate that the onset of drag reduction does vary with mean molecular weight since the molecular weights shown are consistent with the manufacturer specified values.

### 3.2 Degraded bulk flow characterization

It is well documented that when the wall shear rate is sufficiently large mechanical degradation via chain scission is possible [11,32,37]. While a universal scaling law for chain scission based on the molecular bond strength [40] is available, PEO has an established empirical relationship for the shear rate at the onset of degradation ($\gamma_D$) for a given mean molecular weight, $\gamma_D = 3.23 \times 10^{18} M_w^{-2.20}$ [30,63]. If the shear rate exceeds $\gamma_D$, the polymer chains will break and the mean molecular weight will decrease. Within the polymeric regime on a P-K plot, this is realized as data deviating from the logarithmic curve at higher $Re\sqrt{f}$ and bending back towards the P-K law [19,28]. This empirical relationship was used to design the current pressure drop apparatus and select the operation range such that no degradation occurred prior to the pressure drop measurement section. However, downstream of the measurement section was a needle valve that controlled the flowrate, which produced sufficiently high shear rates to rapidly degrade PEO via chain scission (i.e. breaking of the carbon-carbon and carbon-oxygen bonds that make up the polymer backbone). Thus, mechanically degraded samples were produced by passing a sample through the pressure drop apparatus with the needle valve in a predetermined position prior to passing them through a second time to characterize the degraded samples.

An example of a characterization of a degraded sample from the current study is provided in Figure 4. Here a sample with an initial molecular weight $M_{wi} = 2 \times 10^6$ g/mol was degraded to $M_{wf} = 0.6 \times 10^6$ g/mol. For comparison, results from a non-degraded $M_w = 2 \times 10^6$ g/mol sample are also provided along with the P-K law (Eq. (1)) and the MDR asymptote (Eq. (2)). The impact of mechanical degradation on the polymer behavior is apparent from the onset of drag reduction for the degraded sample shifted to the right (i.e. to high Reynolds numbers and shear rates) compared to the non-degraded sample. This is consistent with Vanapalli et al. [30,40] that the lower the mean molecular weight the higher the shear rate at the onset of drag reduction. A summary of the degraded results is provided in Table 3, which includes the initial molecular weight ($M_{wi}$) and final molecular weight ($M_{wf}$) as well as the polymeric regime characterization parameters.

## 4 Discussion and Analysis

### *4.1 Drag reduction performance*

PDR generally is defined based on the reduction of the wall shear stress relative to the Newtonian (e.g. water) flow. Consequently, the common drag reduction (*DR*) definition is based on a comparison between the polymer modified wall shear stress ($\tau_p$) and the Newtonian wall shear stress ($\tau_N$), $DR = (\tau_N - \tau_p)/\tau_N$. As previously observed in the P-K plots, the drag reduction is dependent on the solution concentration, molecular weight and Reynolds number. While these drag reduction levels can be iteratively computed for every test condition, it is more efficient to quantify the drag reduction efficiency of the polymer solution via comparison of the change in the slope relative to the P-K law (i.e. the slope increment, $\delta$). Thus to quantify the impact of mechanical degradation on the drag reduction ability of the polymer, degraded and non-degraded samples with the same onset of drag reduction (i.e. nominal mean molecular weight) and concentration were characterized and their resulting slope increments compared. The degradation procedure was tuned such that PEO samples with a nominal initial molecular weight between $1 \times 10^6$ and $5 \times 10^6$ g/mol were degraded to match the onset of drag reduction condition (i.e. mean molecular weight) as the non-degraded samples, which ranged from $0.6 \times 10^6$ to $4 \times 10^6$ g/mol.

Degraded samples are compared with non-degraded samples at the same onset of drag reduction condition (i.e. mean molecular weight) in Figure 5. First, the examination of the samples that had $M_{wi} = 4 \times 10^6$ g/mol that were degraded to $M_{wf} = 2 \times 10^6$ g/mol with either $C = 100$ ppm or 500 ppm shows excellent agreement between the degraded and non-degraded samples. The maximum relative difference between degraded and non-degraded samples was ~5%, which is within the uncertainty of these measurements. However, the deviation does appear to increase slightly with increasing concentration with the degraded sample consistently having the slightly larger slope, which suggests that the slope increment could have a potential weak concentration dependence. These results indicate that within this range of conditions tested, mechanical degradation has a negligible impact on the drag reduction performance as long as the mean molecular weight (i.e. onset of drag reduction) was matched between samples. Based on observations from Gampert and Wagner [38], this is because, within the range of Reynolds numbers tested *(Re < 35,000)*, the longer chains do not show any preferential extensions over other comparable but slightly shorter chains. This difference in size of chain is not significant enough to drastically change the flow characteristics of the solution and, therefore, no significant difference in bulk behavior was observed.

However, Figure 5 also includes a $C = 500$ ppm sample that had $M_{wi} = 2 \times 10^6$ g/mol that was degraded to $M_{wf} = 0.6 \times 10^6$ g/mol compared with a non-degraded $M_w = 0.6 \times 10^6$ g/mol sample. These samples reveal a significant difference in slope increment between the mechanical degraded and non-degraded samples. Once again, the degraded sample produces a larger slope increment (i.e. more efficient at reducing the drag). Since the concentration was matched with one of the $M_{wi} = 4 \times 10^6$ g/mol conditions, these results indicate that there is a dependence on the initial and/or final molecular weights of the polymer solution in this regime of original, non-degraded (starting) molecular weights. This dependence on molecular weights indicates that the magnitude of drag reduction rather strongly depends on the largest molecular weight chains present in the solution. Gampert and Wagner [38] showed that a Reynolds number of 20,000 was enough to degrade the fractions of large chains. Since the Reynolds number range of these experiments are above 20,000, the observations made with the mentioned molecular weight regimes points to the active participation of the long chains because they are stretched preferentially within the range of shear fields imposed by the Reynolds number. It is worth mentioning that in general these deviations were not strongly dependent on polymer concentration for a given molecular weight, which is convenient since the concentration range was dependent on the sample molecular weight given the flow setup (i.e. if concentration needed to be matched for all conditions, separate pressure drop apparatuses would have to be made for various molecular weight ranges).

Additional tests using non-degraded or initial molecular weights of 0.6, 1.0, 2.0 ,4.0 and $5.0 \times 10^6$ g/mol showed similar trends. While slope increment is relatively independent of the pipe diameter, it is sensitive to the polymer concentration, polymer-solvent combination and molecular weight [60]. This prevents a simple comparison of variation in slope increment between degraded and non-degraded samples since the polymer concentration could not be matched for all combinations (i.e. higher molecular weight samples required lower concentrations than lower molecular weight samples). However, Virk [60] compared numerous combinations of polymer types and solvents and showed that the slope increment is well approximated as being proportional to the square root of concentration. This relationship is shown in Figure 6 with all the best fits curves having slopes of 0.5. Note that a power-law fit to the raw data produces exponents that were ±5% of 0.5 for all conditions. These results also show the sensitivity of the slope increment to molecular weight as well as the discrepancy observed in Figure 5 between the degraded and non-degraded $M_w = 0.6 \times 10^6$.

If the molecular weight (i.e. onset of drag reduction) and the polymer-solvent (PEO-water) are matched, the degraded and non-degraded samples should produce the same $\delta$ versus $C$ curves if the drag reduction performance has been unaltered. Thus a robust means of quantifying the deviation between the degraded and non-degraded samples is to examine the difference in the slope increment for a given concentration. Based on the previous observations, the deviation observed between the slope increment of degraded ($\delta_D$) and non-degraded ($\delta_{ND}$) samples appeared to be dependent on the initial and/or final molecular weight of the samples. Since the slope increment deviation should approach zero as degradation approaches zero, a reasonable parameter to scale the deviation is the difference between the initial and final molecular weights. This difference was normalized with initial molecular weight to make the scaling parameter, $\zeta = (M_{wi} - M_{wf})/M_{wi}$. Figure 7 shows the deviation of the slope increment ($\delta_D - \delta_{ND}$) as a function of $\zeta$. The first observation from these results is that for all conditions tested the slope increment of the degraded samples was higher than the non-degraded samples at the same mean molecular weight. In addition, these results show a relatively small deviation for $\zeta < 0.6$, followed by a rapid increase in the deviation.

## *4.2 Polydispersity*

Since testing was performed within the same flow operation range (Reynolds number, geometry, etc), the deviation must be the product of variations within polymer properties. Most PEO polymer properties (e.g. relaxation time, viscosity ratio, length ratio) are primarily a function of the molecular weight and concentration. Since the concentration and mean molecular weight (i.e. onset of drag reduction shear rate) are equal between the degraded and non-degraded PEO samples, the deviation in performance within the polymeric regime must be related to variations in the distribution of the molecular weight. This assessment is supported by Paterson and Abernathy [31] that determined that the post-degradation molecular weight distribution (e.g. polydispersity) is critical to interpreting polymer flow properties. Furthermore, Hunston and Zakin [37] studied the effect of mechanical degradation (flow-assisted degradation) on the molecular weight distribution of polystyrene. Turbulent measurements, like those in the current study, were used to broaden the range of conditions that could be studied with viscosity or GPC methods. This work showed that for polystyrene the onset of drag reduction was dependent on the molecular weight with the results biased towards the largest molecules in the sample, and that flow rate dependence was related to the shape of the top part of the molecular weight distribution. This supports the assessment that the deviations in slope increment with the PEO samples were likely the product of a change in the molecular distribution (polydispersity) of the samples.

While this suggests that the deviations are related to changes in the molecular weight distribution, it does not explain the consistent improvement of the degraded samples relative to the non-degraded (at the same mean molecular weight) samples. The longer chain molecules have the greatest impact on determining the flow properties of a solution due to their preferential mode of extension [38], which suggests that the current samples (especially those with the largest $\zeta$) had a larger percentage of the longer chain molecules than the non-degraded samples. In general, mechanical degradation narrows the molecular weight distribution if the shear rate is relatively uniformly applied [32,49]. Wall-bounded flows (e.g. pipes, boundary layers) do not have uniformly applied shear rates, which results in a relatively small percentage of the chains being stretched to lengths comparable to the polymer contour length (i.e. maximum polymer extension length) at any instance in time [61,62]. However, if the polymer chains are exposed to the turbulent wall-bounded flow for a sufficiently long period of time a steady-state condition can be achieved once a sufficient number of stretching/degradation cycles are achieved [11]. If the elongational rate far exceeds that of the critical elongational rate, then the midpoint scission assumption [33-35] would be violated and the final (steady-state) distribution would be broader than the initial [36]. Prior to achieving steady-state conditions, the molecular weight distribution would be asymmetric and biased towards higher molecular weights because at each time step some percentage of chains would not have broken yet. This suggests that the current results correspond to an intermediate stage of degradation (i.e. prior to achieving steady-state behavior), which was confirmed by comparing results after multiple passes through the pressure drop apparatus.

The deviation in slope increment for $\zeta > 0.6$ is also indicative of the fact that a mere presence of a few long chain polymer molecules within a solution can be responsible for significantly increasing the drag reduction. That is to say, these small fractions of long chain molecules have a greater impact in defining the flow properties of a degraded sample, than the mean molecular weight of the sample [31]. The validity of this claim within the specified regime of $\zeta$, is also subject to the Reynolds number range tested, which for the current study was rarely above 30,000. For this range of Reynolds numbers, it could be justified to say that the long chain polymers show preferential extension over the shorter chains and therefore control flow properties of the solutions. Such a behavior is expected to be more pronounced when the disparity between short chains and long chains within a solution is large (disparity in terms of their molecular weight averages). Although Gampert and Wagner [38] used artificially created polydispersed synthesized PAM solutions, they reached the same conclusions, which provides additional support to the validity of these conclusions.

The functional relationship for the $\zeta$ dependence as shown in Figure 7, and more specifically the value where significant variation was observed, is most likely specific to the degradation process. If the residence time were increased, it is presumed that a larger value of $\zeta$ could be achieved without significant deviations in the slope increment since any variation would be the product of the broadening of the distribution [36] rather than an excess of larger molecules. As the ratio of the residence time to relaxation time becomes large, the steady-state molecular weight would be achieved and the impact of $\zeta$ is expected to significantly decrease if these assumptions are valid. This was tested by creating a PEO polymer ocean with $C = 100$ ppm within the Oklahoma State University 6-inch low-turbulence, recirculating water tunnel [64]. This allowed the facility to be operated for as long as was required to achieve a steady-state mean molecular weight (based on the onset of drag reduction). In addition, the speeds were selected so that the steady-state molecular weights matched two of the non-degraded molecular weights ($M_{wf}$ of $0.6\times10^6$ g/mol and $2\times10^6$ g/mol). The results for the steady-state degraded samples are shown in Figure 8 along with their corresponding non-degraded samples. The deviations in the slope increment for the $0.6\times10^6$ g/mol and $2\times10^6$ g/mol samples were $\delta_D - \delta_{ND} < 0.5$ ($\zeta = 0.8$) and $\delta_D - \delta_{ND} = 1.3$ ($\zeta = 0.5$), respectively. These variations are within the measurement uncertainty and illustrate the difference from that observed in Figure 7, which supports the conjecture that these deviations can be mitigated if steady-state conditions can be achieved.

*Mechanical Degradation of PEO*　　　　　　　　　　　　　　　　　　　　*Farsiani, Saeed and Elbing*# 5  Conclusion

The current study uses a turbulent pipe flow experiment to do a comparative analysis between mechanically degraded polymer (PEO) solutions and non-degraded polymer (PEO) solutions at the same mean molecular weight. Degraded samples were produced via passing samples through a pipe that included a precisely positioned V-shaped needle valve. The degradation resulted in an increase in the shear rates at the onset of drag reduction, which Vanapalli et al. [30] provided an empirical relationship between the onset of drag reduction shear rate and the mean molecular weight for PEO. The samples were degraded such that they produced mean molecular weights (onset of drag reduction shear rates) that matched available non-degraded molecular weights. Characterization of the non-degraded samples produced bulk flow behavior that is consistent with previous PEO studies in the literature [28,30,38,60]. Comparative analysis of the mechanically degraded samples (samples with different initial, but known, mean molecular weights degraded to a known final mean molecular weight) with the non-degraded samples at the mean molecular weight of the final state of the degraded samples produced the following primary conclusions:

1) While some conditions showed good agreement in the slope increment between degraded and non-degraded samples (Figure 5), there were conditions that had significant deviations in the slope increment (drag reduction performance) between the degraded and non-degraded samples, with the non-degraded samples consistently larger (more efficient).

2) The deviation in slope increment scaled well with the normalized difference between the initial and final molecular weights, $\zeta = (M_{wi} - M_{wf})/M_{wi}$ with the deviation increasing rapidly when $\zeta > 0.6$. However, it is expected that the exact value of this acceleration is specific to the degradation method, including the ratio of the residence time to the relaxation time.

3) The deviations in drag reduction performance for degraded and non-degraded samples at a given molecular weight were attributed to deviations in the molecular weight distribution, which was supported by other observations in the literature [31,37]. Furthermore, this behavior is likely enhanced prior to achieving steady-state molecular weight when there would be an excess of longer polymer chains [38], which was the case for the majority of conditions presented.

4) The amount of deviation can be reduced if steady-state conditions can be achieved. However, if the elongational rate far exceeds the critical elongational rate then the final molecular weight distribution could be broader [36], which could still impact the drag reduction performance.

These results provide criteria that should be followed if comparisons in drag reduction performance will be made between mechanically (flow-assisted) degraded and non-degraded samples. These results are particularly valuable when using high molecular weight PEO samples at relatively low concentrations (i.e. common drag reduction operation conditions) since the common viscosity and GPC approaches are not well suited for these conditions [38,58]. This also enables a robust means of establishing polymeric oceans that can be compared with previous non-degraded samples, which can greatly simplify fundamental PDR studies of developing turbulent boundary layers by removing the concentration dependence.

# Acknowledgments

These authors would like to thank Dr. Frank Blum and Dr. Siva Vanapalli for discussions on the use of GPC for analysis of high-molecular weight PEO as well as Marcus Lander, Jacquelyne Baade and JW Wallace who have assisted with experiments.

# Funding

This work was supported by the National Science Foundation under Grant No. 1604978 (Dr. Ronald Joslin, Program Manager).

# References

[1] Mysels KJ (1947) Flow of thickened fluids. U.S. Patent 2,492,173, priority date June 12, 1947.

[2] Toms BA (1948) Some observations on the flow of linear polymer solutions through straight tubes at large Reynolds numbers. In: Proceedings of the 1$^{st}$ International Congress on Rheology, 2, 135, North-Holland, Amsterdam.

[3] Toms BA (1949) Detection of a wall effect in laminar flow of solutions of a linear polymer. Journal of Colloid Science, 4(5), 511-521.

[4] Hoyt JW (1972) The effect of additives on fluid friction. Journal of Fluids Engineering, 94(2), 258-285.

[5] Burger ED, Chorn LG, Perkins TK (1980) Studies of drag reduction conducted over a broad range of pipeline conditions when flowing Prudhoe Bay crude oil. Journal of Rheology, 24(5), 603-626.

[6] Sellin R, Hoyt J, Poliert J, Scrivener O (1982) The effect of drag reducing additives on fluid flows and their industrial applications part 2: Present applications and future proposals. Journal of Hydraulic Research, 20(3), 235-292.

[7] Fruman DH, Aflalo SS (1989) Tip vortex cavitation inhibition by drag-reducing polymer solutions. Journal of Fluids Engineering, 111(2), 211-216.

[8] White CM, Mungal MG (2008) Mechanics and prediction of turbulent drag reduction with polymer additives. Annual Review of Fluid Mechanics, 40, 235-256.

[9] Elbing BR, Dowling DR, Perlin M, Ceccio SL (2010) Diffusion of drag-reducing polymer solutions within a rough-walled turbulent boundary layer. Physics of Fluids, 22(4), 045102.

[10] Elbing BR, Winkel ES, Ceccio SL, Perlin M, Dowling DR (2010) High-Reynolds-number turbulent-boundary-layer wall-pressure fluctuations with dilute polymer solutions. Physics of Fluids, 22(8), 085104 (doi.org/10.1063/1.3478982).

[11] Elbing BR, Solomon MJ, Perlin M, Dowling DR, Ceccio SL (2011) Flow-induced degradation of drag-reducing polymer solutions within a high-Reynolds-number turbulent boundary layer. Journal of Fluid Mechanics, 670, 337-364.

[12] Elbing BR, Perlin M, Dowling DR, Ceccio SL (2013) Modification of the mean near-wall velocity profile of a high-Reynolds number turbulent boundary layer with the injection of drag-reducing polymer solutions. Physics of Fluids, 25(8), 085103.

[13] Perlin M, Dowling DR, Ceccio SL (2016) Freeman scholar review: Passive and active skin-friction drag reduction in turbulent boundary layers. Journal of Fluids Engineering, 138(9), 091104.

[14] Elbing BR (2018) Flow-assisted polymer degradation in turbulent boundary layers. In: Proceedings of the AIChE Annual Meeting, Area 01C Interfacial Phenomena, Pittsburgh, PA (Oct 28 – Nov 2).

[15] McGary, Jr CW (1960) Degradation of poly(ethylene oxide). Journal of Polymer Sciences, 46 51-57.

[16] Shin H (1965) Reduction of Drag in Turbulence by Dilute Polymer Solutions. Phd Dissertation, Massachusetts Institute of Technology, Cambridge, MA, USA.

[17] Bailey Jr FE, Koleske, JV (1976) Polyethylene Oxide, Academic Press, New York.

[18] Bortel E, Lamot R (1977) Examination of the breakdown of high molecular weight polyethylene oxides in the solid state. Macromolecular Chemistry and Physics, 178(9), 2617-2628.

[19] Moussa T, Tiu C (1994) Factors affecting polymer degradation in turbulent pipe flow. Chemical Engineering Science, 49(10), 1681-1692.

[20] Fore RS, Szwalek J, Sirviente A (2005) The effects of polymer solution preparation and injection on drag reduction. Journal of Fluids Engineering, 127(3), 536-549.

[21] Layec-Raphalen MN, Layec Y (1985) Influence of molecular parameters on laminar non-Newtonian and on turbulent flows of dilute polymer solutions. In: The Influence of Polymer Additives on Velocity and Temperature Fields, International Union of Theoretical and Applied Mechanics (Deutsche Rheologische Gasellschaft) (ed. Gampert B), Springer, Berlin, Heidelberg, 89-100.

[22] Yarin AL (1991) Strong flows of polymeric liquids: Part 2. Mechanical degradation of macromolecules. Journal of Non-Newtonian Fluid Mechanics, 38(2-3), 127-136.

[23] Yarin AL (1993) Free Liquid Jets and Films: Hydrodynamics and Rheology, Longman Scientific & Technical and Wiley & Sons, Harlow, New York.

[24] Yarin AL (1997) On the mechanism of turbulent drag reduction in dilute polymer solutions: Dynamics of vortex filaments. Journal of Non-Newtonian Fluid Mechanics, 69(2-3), 137-153.

[25] Zaitoun A, Makakou P, Blin N, Al-Maamari RS, Al-Hashmi A-AR, Abdel-Goad M (2012) Shear stability of EOR polymers. SPE Journal, 17(2), 335-339.

[26] Fontaine A, Petrie H, Brungart T (1992) Velocity profile statistics in a turbulent boundary layer with slot-injected polymer. Journal of Fluid Mechanics, 238, 435-466.

[27] Petrie H, Fontaine A, Moeny M, Deutsch S (2005) Experimental study of slot-injected polymer drag reduction. In: Proceedings of the 2nd International Symposium on Seawater Drag Reduction, Busan, Korea (May 23-26).

[28] Elbing BR, Winkel ES, Solomon MJ, Ceccio SL (2009) Degradation of homogeneous polymer solutions in high shear turbulent pipe flow. Experiments in Fluids, 47(6), 1033-1044.

[29] Culter JD, Zakin JL, Patterson GK (1975) Mechanical degradation of dilute solutions of high polymers in capillary tube flow. Journal of Applied Polymer Science, 19(12), 3235-3240.

[30] Vanapalli SA, Islam MT, Solomon MJ (2005) Scission-induced bounds on maximum polymer drag reduction in turbulent flow. Physics of Fluids, 17(9), 095108.

[31] Paterson RW, Abernathy FH (1970) Turbulent flow drag reduction and degradation with dilute polymer solutions. Journal of Fluid Mechanics, 43(4), 689-710.

[32] Yu JFS, Zakin JL, Patterson GK (1979) Mechanical degradation of high molecular weight polymers in dilute solution. Journal of Applied Polymer Science, 23(8), 2493-2512.

[33] Hinch EJ (1977) Mechanical models of dilute polymer solutions in strong flows. The Physics of Fluids, 20(10), S22-S30.

[34] Horn A, Merrill E (1984) Midpoint scission of macromolecules in dilute solution in turbulent flow. Nature, 312(5990), 140.

[35] Odell JA, Keller A, Miles MJ (1983) Method for studying flow-induced polymer degradation: verification of chain halving. Polymer Communications, 24(1), 7-10.

[36] Sim H, Khomami B, Sureshkumar R (2007) Flow–induced chain scission in dilute polymer solutions: Algorithm development and results for scission dynamics in elongational flow. Journal of Rheology, 51(6), 1223-1251.

[37] Hunston DL, Zakin JL (1980) Flow-assisted degradation in dilute polystyrene solutions. Polymer Engineering and Science, 20(7), 517-523.

[38] Gampert B, Wagner P (1985) The Influence of molecular weight and molecular weight distribution on drag reduction and mechanical degradation in turbulent flows of highly dilute polymer solutions. In: The Influence of Polymer Additives on Velocity and Temperature Fields, International Union of Theoretical and Applied Mechanics (Deutsche Rheologische Gasellschaft) (ed. Gampert B), Springer, Berlin, Heidelberg, 71-85.

[39] Habibpour M, Clark PE (2017) Drag reduction behavior of hydrolyzed polyacrylamide/xanthan gum mixed polymer solutions. Petroleum Science, 14(2), 412-423.

[40] Vanapalli SA, Ceccio SL, Solomon, MJ (2006) Universal scaling for polymer chain scission in turbulence. Proceedings of the National Academy of Sciences, 103(45), 16660-16665.

[41] Grandbois M, Beyer M, Rief M, Clausen-Schaumann H, Gaub HE (1999) How strong is a covalent bond? Science, 283(5408), 1727-1730.

[42] White CM, Somandepalli VSR, Mungal MG (2004) The turbulence structure of drag-reduced boundary layer flow. Experiments in Fluids, 36(1), 62-69.

[43] Hou YX, Somandepalli VSR, Mungal MG (2008) Streamwise development of turbulent boundary-layer drag reduction with polymer injection. Journal of Fluid Mechanics, 597, 31-66.

[44] Somandepalli VSR, Hou YX, Mungal MG (2010) Concentration flux measurements in a polymer drag-reduced turbulent boundary layer. Journal of Fluid Mechanics, 644, 281-319.

[45] White CM, Dubief Y, Klewicki J (2012) Re-examining the logarithmic dependence of the mean velocity distribution in polymer drag reduced wall-bounded flow. Physics of Fluids, 24(2), 021701.

[46] Escudier M, Rosa S, Poole R (2009) Asymmetry in transitional pipe flow of drag-reducing polymer solutions. Journal of Non-Newtonian Fluid Mechanics, 161(1-3), 19-29.

[47] Farsiani Y, Elbing BR (2017) Modification of a turbulent boundary layer within a homogeneous concentration of drag reducing polymer solution. APS Division of Fluid Dynamics Annual Meeting, Q28.5, Denver, CO (Nov. 19-21); In: Bulletin of the American Physical Society, 62(14).

[48] Farsiani Y, Elbing BR (2018) Effect of flow and polymer properties on near wall mean and fluctuating velocity profiles. APS Division of Fluid Dynamics Annual Meeting, A17.4, Atlanta, GA (Nov. 18-20); In: Bulletin of the American Physical Society, 63(13).

[49] Kim CA, Kim JT, Lee K, Choi HJ, Jhon MS (2000) Mechanical degradation of dilute polymer solutions under turbulent flow. Polymer, 41(21), 7611-7615.

[50] Kalashnikov VN (2002) Degradation accompanying turbulent drag reduction by polymer additives. Journal of Non-Newtonian Fluid Mechanics, 103(2-3), 105-121.

[51] Draad AA, Kuiken G, Nieuwstadt F (1998) Laminar–turbulent transition in pipe flow for Newtonian and non-Newtonian fluids. Journal of Fluid Mechanics, 377, 267-312.

[52] Petrie HL, Deutsch S, Brungart TA, Fontaine AA (2003) Polymer drag reduction with surface roughness in flat-plate turbulent boundary layer flow. Experiments in Fluids, 35(1), 8-23.

[53] De Gennes P.-G (1979) Scaling concepts in polymer physics, Cornell University Press, Ithaca.

[54] Bailey Jr FE, Callard RW (1959) Some properties of poly(ethylene oxide) in aqueous solution. Journal of Applied Polymer Science, 1(1), 56-62.

[55] Lander M (2018) Preparation and characterization of polyethylene-oxide (PEO) solution. MS Thesis, Oklahoma State University, Stillwater, USA.

[56] McKeon BJ (2007) Measurement of pressure with wall tappings. In: Handbook of Experimental Fluid Mechanics (ed. Tropea C, Yarin AL, Foss JF), Springer, 180-184.

[57] Kulik VM (2001) Drag reduction change of polyethyleneoxide solutions in pipe flow. Experiments in Fluids, 31(5), 558-566.

[58] Berman NS (1977) Drag reduction of the highest molecular weight fractions of polyethylene oxide. Physics of Fluids, 20(5) 715-718.

[59] Virk PS, Merrill EW, Mickle HS, Smith KA, Mollo-Christensen EL (1967) The Toms phenomenon: turbulent pipe flow of dilute polymer solutions. Journal of Fluid Mechanics, 30(2), 305-328.

[60] Virk PS (1975) Drag reduction fundamentals. AIChE Journal, 21(4), 625-656.

[61] Dubief Y, White CM, Terrapon VE, Shaqfeh ES, Moin P, Lele SK (2004) On the coherent drag-reducing and turbulence-enhancing behavior of polymers in wall flows. Journal of Fluid Mechanics, 514, 271-280.

[62] Gupta V, Sureshkumar R, Khomami B (2004) Polymer chain dynamics in Newtonian and viscoelastic turbulent channel flows. Physics of Fluids, 16(5), 1546-1566.

[63] Winkel ES, Oweis GF, Vanapalli SA, Dowling DR, Perlin M, Solomon MJ, Ceccio SL (2009) High-Reynolds-number turbulent boundary layer friction drag reduction from wall-injected polymer solutions. Journal of Fluid Mechanics, 621, 259-288.

[64] Elbing BR, Daniel L, Farsiani Y, Petrin CE (2018) Design and validation of a recirculating, high-Reynolds number water tunnel. Journal of Fluids Engineering, 140(8), 081102.

# Figures and Tables

Table 1. Summary of the range of molecular weights and concentrations tested in the current study as well as the corresponding intrinsic viscosity and overlap concentration for the given molecular weight PEO.

| $M_w \times 10^{-6}$ (g/mol) | $C$ range (ppm) | $[\eta]_0$ (cm$^3$/g) | $C^*$ (ppm) |
|---|---|---|---|
| 0.6 | 100 − 500 | 402 | 2500 |
| 1 | 500 | 598 | 1680 |
| 2 | 50 − 500 | 1030 | 975 |
| 4 | 5 − 100 | 1760 | 568 |
| 5 | 5 − 20 | 2100 | 477 |

Table 2. Summary of non-degraded samples tested in the pressure drop apparatus as well as the resulting slope increment ($\delta$), onset wave number ($W^*$) and the shear rate at the onset of drag reduction ($\gamma^*$).

| $M_w \times 10^{-6}$ (g/mol) | $C$ range (ppm) | $\delta$ (--) | $W^*$ at $C_{max}$ (m$^{-1}$) | $\gamma^*$ (s$^{-1}$) |
|---|---|---|---|---|
| 0.6 | 100 − 500 | 3.1 − 7.5 | 85,200 | 6,090 |
| 1 | 500 | 13.5 | 71,900 | 3,050 |
| 2 | 50 − 500 | 5.2 − 14.5 | 47,700 | 1,950 |
| 4 | 5 | 6.13 | 30,900 | 900 |
| 5 | 5 | 7.5 | 27,800 | 697 |

Table 3. Summary of degraded PEO samples tested in the pressure drop apparatus.

| $Mw_i \times 10^{-6}$ (g/mol) | $Mw_f \times 10^{-6}$ (g/mol) | $C$ (wppm) | $Re \times 10^{-3}$ at $C_{max}$ range (--) | $\delta$ (--) | $\gamma^*$ (1/s) | $\gamma \times 10^{-3}$ range (1/s) |
|---|---|---|---|---|---|---|
| 5.0 | 1.0 | 500 | 15 − 25 | 22.5 | 2800 | 4.2 − 8.1 |
| 4.0 | 1.0 | 500 | 15 − 30 | 20.6 | 3000 | 4.5 − 12 |
| 4.0 | 2.0 | 50 − 500 | 12 − 28 | 4.5 − 11.0 | 1600 | 1.8 − 10 |
| 5.0 | 2.0 | 500 | 12 − 24 | 16.6 | 2000 | 3.3 − 7.6 |
| 5.0 | 4.0 | 5 | 8 − 24 | 6.1 | 900 | 1.2 − 8.8 |
| 2.0 | 0.6 | 200 − 500 | 18 − 31 | 3.5 − 10.1 | 6090 | 7.7 − 15 |

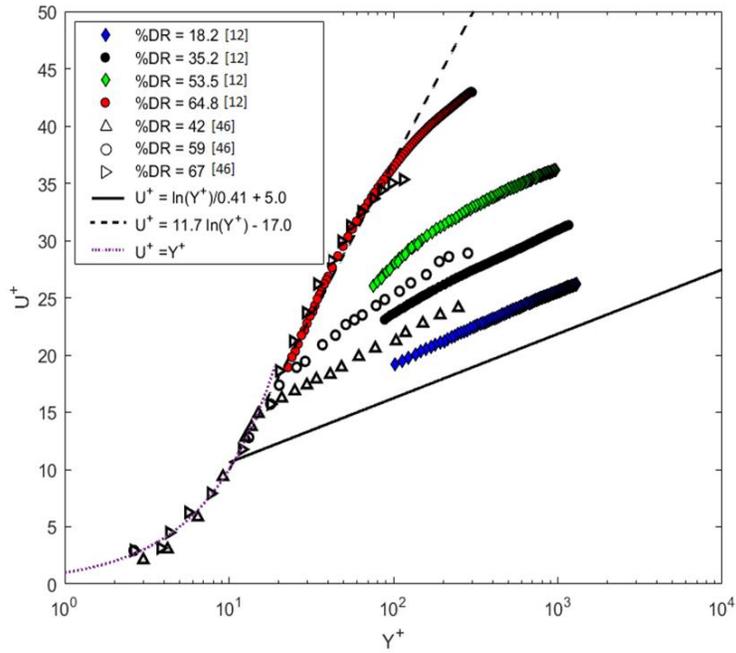

Figure 1. Polymeric velocity profiles from a turbulent boundary layer with non-uniform concentration distribution [12] and channel flow with a constant concentration [46], which shows an increasing slope with increasing drag reduction for HDR (DR > 40%).

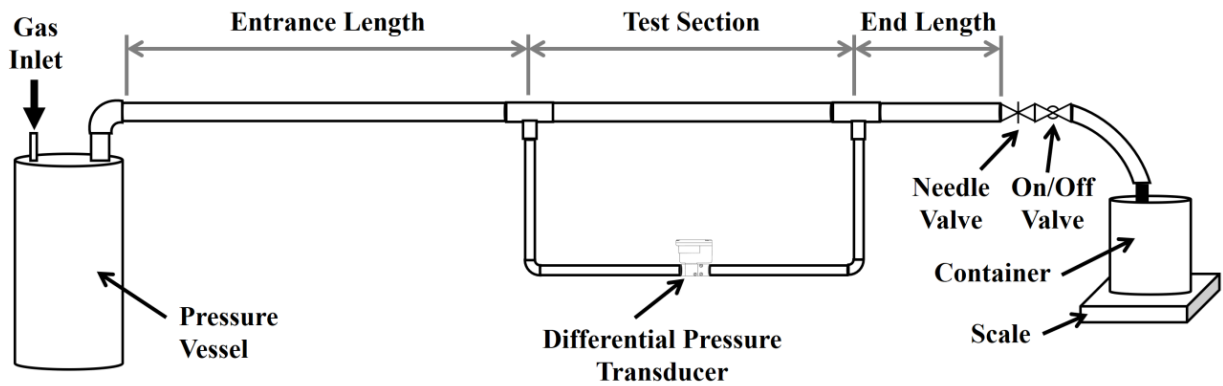

Figure 2. Schematic of the pressure drop apparatus used for characterization of the polymer samples as well as mechanically degrading samples.

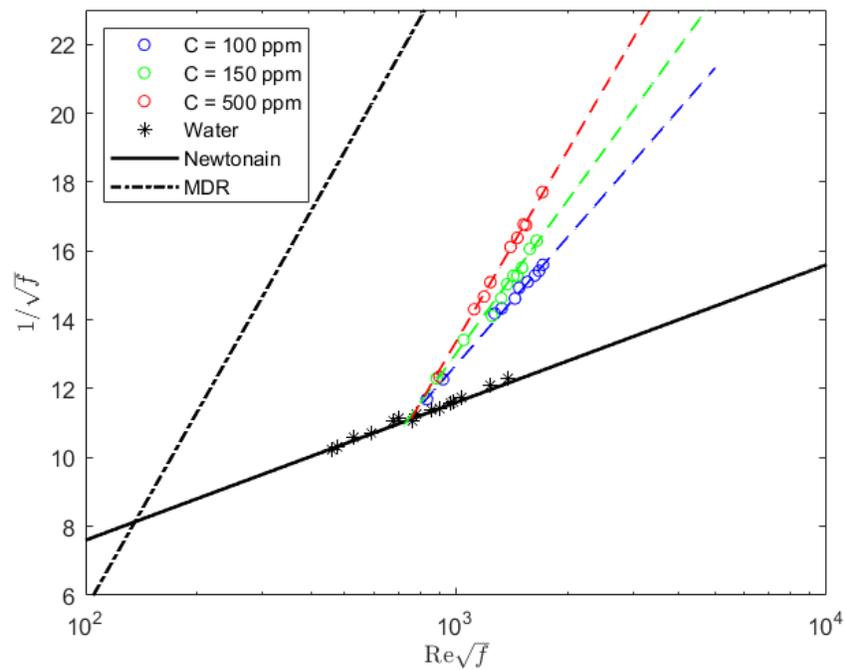

Figure 3. P-K plot of $2\times10^6$ g/mol PEO at concentrations of 100, 150 and 500 ppm, as well as water (Newtonian) data at the same range of $Re\sqrt{f}$. Included for reference are the P-K law, MDR asymptote and logarithmic best-fit curves to the data within the polymeric regime.

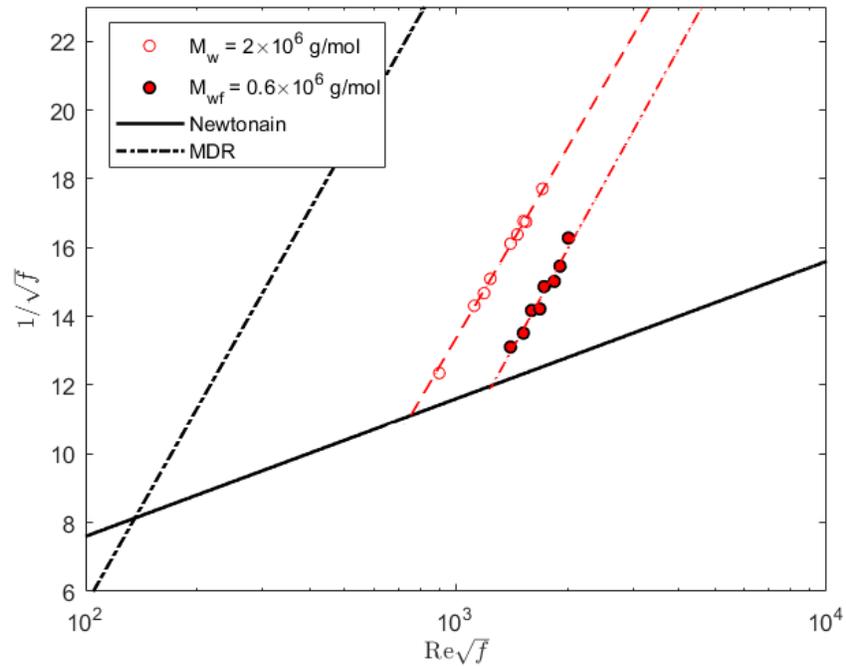

Figure 4. P-K plot using PEO at an initial $M_w = 2\times10^6$ g/mol and $c = 500$ ppm. One of the samples was degraded to a lower molecular weight ($M_w = 0.6\times10^6$ g/mol) while the other was non-degraded.

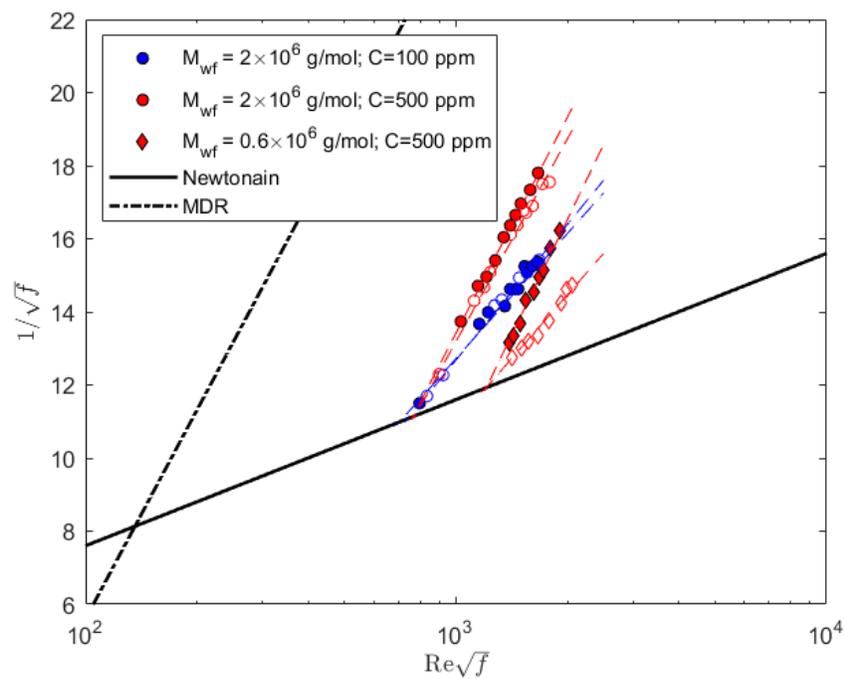

Figure 5. P-K plot comparing degraded and non-degraded samples with $M_{wf} = 2\times10^6$ g/mol or $0.6\times10^6$ g/mol. Filled markers represent degraded samples.

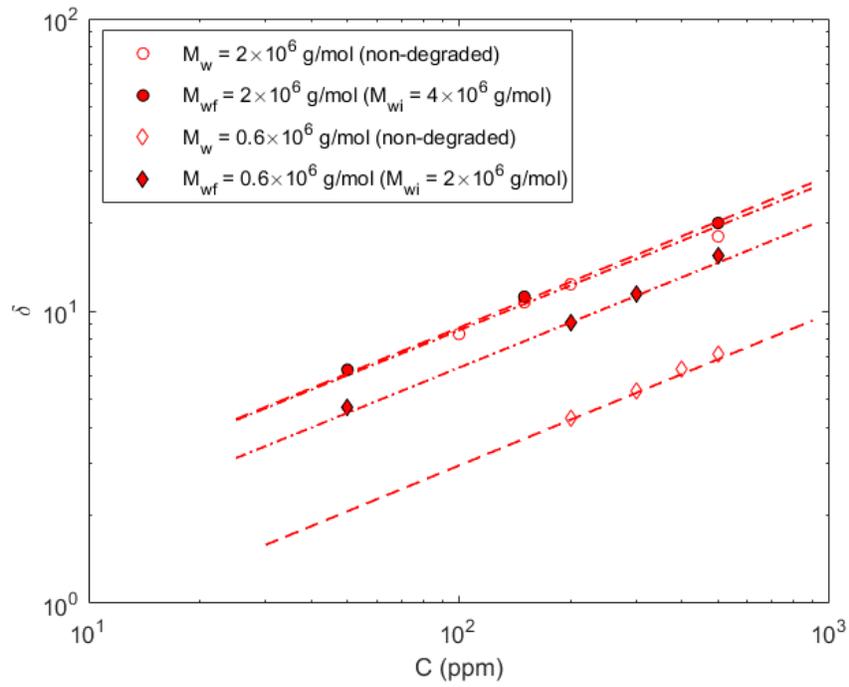

Figure 6. Slope increment versus concentration (*C*) for degraded and non-degraded samples of $M_w = 0.6 \times 10^6$ or $2.0 \times 10^6$ g/mol. The dashed lines are all best fit curves with a slope of 0.5.

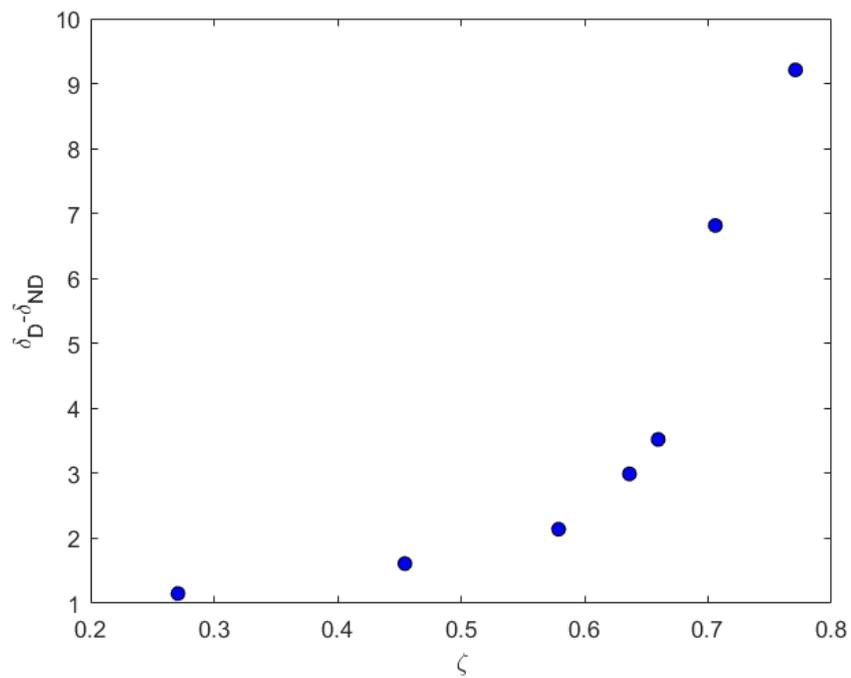

Figure 7. The difference between the degraded and non-degraded slope increments plotted versus the normalized difference between the initial and final molecular weights, $\zeta = (M_{wi} - M_{wf})/M_{wi}$.

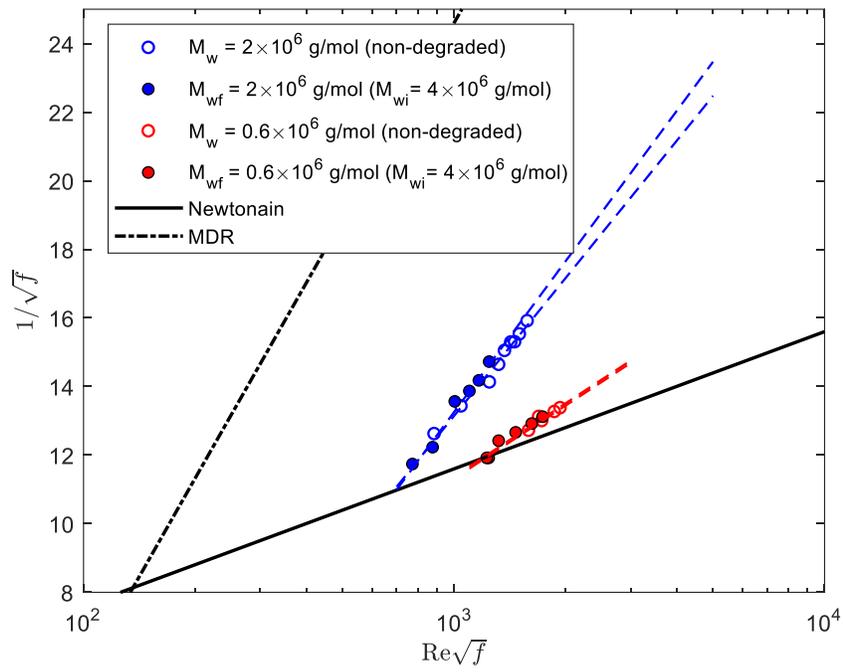

Figure 8. P-K plot comparing a steady-state degraded sample from a $C$ = 100 ppm PEO polymer ocean in a recirculating water tunnel with that of a non-degraded sample.